**Testing the Empirical Shock Arrival Model using Quadrature Observations**

N. Gopalswamy[1], P. Mäkelä[1,2], H. Xie[1,2], and S. Yashiro[1,2]
[1]Code 671, NASA Goddard Space Flight Center, Greenbelt, MD 20771, USA
[2]Department of Physics, The Catholic University of America, Washington, DC 20064, USA

Abstract: The empirical shock arrival (ESA) model was developed based on quadrature data from *Helios* (in-situ) and *P-78* (remote-sensing) to predict the Sun-Earth travel time of coronal mass ejections (CMEs) [*Gopalswamy et al*. 2005a]. The ESA model requires earthward CME speed as input, which is not directly measurable from coronagraphs along the Sun-Earth line. The Solar Terrestrial Relations Observatory (STEREO) and the Solar and Heliospheric Observatory (SOHO) were in quadrature during 2010 - 2012, so the speeds of Earth-directed CMEs were observed with minimal projection effects. We identified a set of 20 full halo CMEs in the field of view of SOHO that were also observed in quadrature by STEREO. We used the earthward speed from STEREO measurements as input to the ESA model and compared the resulting travel times with the observed ones from L1 monitors. We find that the model predicts the CME travel time within about 7.3 hours, which is similar to the predictions by the ENLIL model. We also find that CME-CME and CME-coronal hole interaction can lead to large deviations from model predictions.

# 1. Introduction

Coronal mass ejections (CMEs) are the cause of the severest of geomagnetic storms and large energetic particle events [see *Gopalswamy*, 2009 and references therein]. CMEs appearing close to the disk center of the Sun often appear as halo CMEs [*Howard et al.*, 1982 *Gopalswamy et al.*, 2010a] and are highly likely to intersect Earth. Unfortunately, it is difficult to measure the nose part of the CMEs arriving at Earth by coronagraphs located along the Sun-Earth line because of the occulting disk. The nose part can be directly measured only by the in-situ instruments in the solar wind. On the other hand, the nose part of the CME can be observed by both remote-sensing and in-situ observations if the observing spacecraft are in quadrature. For example, the Solwind coronagraph on board P-78 mission located in Earth orbit observed limb CMEs heading towards the Helios spacecraft that detected the CMEs in situ [*Lindsay et al.* 1999]. The combined in situ and coronagraphic observations were useful in deriving the interplanetary (IP) acceleration of CMEs, free from projection effects. *Gopalswamy et al.* [2001] obtained the IP acceleration of CMEs using Solwind coronagraph data and Helios in-situ observations reported in *Lindsay et al.* [1999]. The IP acceleration was more accurate than the one derived from coronagraphic images obtained by the Solar and Heliospheric Observatory (SOHO) and in-situ measurements from the Wind spacecraft because both Wind and SOHO were located along the Sun-Earth line. The IP acceleration derived by *Gopalswamy et al.* [2001] is of the form,

$$a = -0.0054 (u - u_c), \qquad (1)$$

where $u$ is the initial speed of the CME in coronagraph field of view (FOV) and $u_c$ = 406 km/s, identified as the average solar wind speed. The acceleration can be positive or negative depending on whether the CME speed is smaller or greater than the solar wind speed. The acceleration in eq. (1) provides a simple means to predict the CME travel time to Earth using simple kinematic relations [*Gopalswamy et al.* 2001]. Note that the acceleration has also been modeled to have quadratic dependence CME relative speed [see *Vrsnak et al.* 2013 for details]. By reducing the CME travel time by the average shock standoff time at 1 AU, *Gopalswamy et al.* [2005a] derived the empirical shock arrival (ESA) model. For initial CME speeds > 450 km/s, the ESA model travel time $T$ (in h) can be approximated by the formula,

$$T = AB^u + C, \qquad (2)$$

with $A$ = 151.002, $B$ = 0.998625, and $C$ = 11.5981 [*Gopalswamy et al.* 2005b]. The ESA model has been tested using sky-plane speed from SOHO correcting for projection effects using a cone model [*Xie et al.* 2006]. Quadrature observations became available in 2010 from SOHO and the twin spacecraft of the Solar Terrestrial Relations Observatory [STEREO, *Kaiser et al.*, 2008] mission, allowing us to directly obtain the earthward speed of CMEs without modeling. Making use of the SOHO-STEREO quadrature observations, we report on the performance of the ESA model for a set of 20 full halo CMEs, whose space speed was measured by one of the two STEREO spacecraft: STEREO-Ahead (STA) or STEREO-behind (STB). The number of events used in this paper is similar to the original number of events (19) observed in quadrature and used for deriving the IP acceleration [*Gopalswamy et al.* 2001].

## 2. Quadrature Observations and ESA Model Output

The primary data we use are coronagraphic images from the Large Angle and Spectrometric Coronagraph [LASCO, *Brueckner et al.*, 1995] on board SOHO and the COR2 coronagraph of the Sun Earth Connection Coronal and Heliospheric Investigation (*SECCHI*) instrument suite on board STEREO [*Howard et al*. 2008]. The SOHO and STEREO spacecraft were within $20^o$ of quadrature from the beginning of 2010 (STB: E71, STA: W65) to the beginning of 2012 (STB: E118; STA: W110). Depending on the heliographic location of the CME source, it was possible to find one of the STEREO spacecraft to be in quadrature with SOHO even though the SOHO-STEREO separation angles were different from $90^o$. Two spacecraft are in quadrature when the two views are orthogonal, but we allow a deviation of ~$30^o$. For example, the 19 January 2012 CME originated from N32E22 in Earth view. On that day, STB was at E113. Therefore, STB viewed the solar source as W91 (just a degree behind the limb), and hence was in quadrature with SOHO. On the other hand, STA was at W108 and viewed the source at E130, which was $40^o$ behind the limb in the STA view. The measurements of this CME made from STB had minimal projection effects, so we used the STB measurements for this event. Table 1 lists the halo CMEs used for testing the ESA model. The selection criteria used are: (i) the CME should be observed as a full halo CME by SOHO (Earth view), (ii) the halo CME speed should be ≥ 450 km/s so that the formula in equation (2) can be used to derive the travel time to Earth, and (iii) the CME should be driving a shock at L1 as detected by SOHO's Charge, Element, and Isotope Analysis System/Mass Time-of-Flight (CELIAS/MTOF) experiment [*Ipavich et al*. 1998]. We used the shock list compiled and made available on line at the SOHO MTOF web site for this study (http://umtof.umd.edu/pm/figs.html).

The list of shocks (date and time), the solar wind speed (in km/s) upstream of the shock, and the driving CMEs at the Sun (date and time) are listed in the first three columns of Table 1. The CME time refers to the first appearance of the CME in the STEREO/COR2 FOV (column 3). The solar source of the CME is identified as the heliographic coordinates of the eruption location observed in EUV images either from the Solar Dynamics Observatory (SDO) or STEREO (column 4). We also cross checked the source location with the flare location listed in the online Solar Geophysical Data. The deviation of the CME source from the sky plane as viewed by STA and STB are given in column 5. The first and second numbers correspond to the angular distance of the solar source from the sky plane for STA and STB, respectively. Positive (negative) numbers indicate that the source is behind (in front of) the limb. These numbers were obtained as the difference between the source location in Earth view and the Earth-Sun-spacecraft angle for STA and STB. This difference angle needs to be smaller than $30^o$ for quadrature. The CME height-time measurements were made by the SOHO/LASCO operator as soon as the halo CME appeared in LASCO FOV using LASCO and STEREO images. Since the CME was a halo in the LASCO FOV, the speed was underestimated. We used the speed measured in that STEREO/COR2 FOV in which the CME was closest to the limb as the true space speed ($V_{sp}$) of the CME.

*Table 1. List of shocks detected at L1 and the corresponding halo CMEs observed by SOHO*

| Shock Date Time | $V_{SW}$ km/s | CME Time | Location | To Limb (A,B) Deg | $V_{Sp}$ km/s S/C | $V_E$ km/s | $t_{ESA}$ h | $t_{obs}$ h | $\Delta t$ h | $V_{E1}$ km/s | $t_{ESA1}$ h | $\Delta t_1$ h | $\Delta t_2$ h |
|---|---|---|---|---|---|---|---|---|---|---|---|---|---|
| 2010/02/15 17:28 | 300 | 02/12 13:31 | N26E11 | -14, -30[b] | 867 B | 765 | 64.3 | 76.0 | -11.7 | 756 | 65.0 | -11.0 | ---- |
| 2010/04/11 12:18 | 360 | 04/08 04:30 | N24E16 | -6, -35 | 771[c] A | 677 | 71.1 | 79.8 | -8.7 | 630 | 75.1 | -4.7 | -3.3 |
| 2010/08/03 16:51 | 425 | 08/01 08:24 | N20E36 | 23, -64 | 1031[c] A | 784 | 62.9 | 56.5 | +6.4 | 1257 | 38.4 | -18.1 | -6.9 |
| 2011/02/18 00:40 | 325 | 02/15 02:36 | S12W18 | -21, 22 | 945 A | 879 | 56.7 | 70.1 | -13.4 | 864 | 57.6 | -12.5 | -9.8 |
| 2011/03/10 05:45 | 300 | 03/07 14:48 | N11E21 | 19, -16 | 691 B | 633 | 74.8 | 63.0 | +11..8 | 738 | 66.3 | +3.3 | 1.8 |
| 2011/06/23 02:18 | 600 | 06/21 03:16 | N16W08 | -2, 11 | 986 A | 939 | 53.1 | 47.0 | +6.1 | 812 | 61.0 | +14.0 | +9.6 |
| 2011/08/04 21:10 | 350 | 08/02 06:36 | N14W15 | -5, 18 | 1015 A | 951 | 52.4 | 62.6 | -10.2 | 883 | 56.4 | -6.2 | ---- |
| 2011/08/05 17:23 | 350 | 08/03 13:17 | N22W30 | -20, 33 | 1322 A | 1062 | 46.6 | 52.1 | -5.5 | 1161 | 42.2 | -9.9 | -0.4 |
| 2011/08/05 18:32 | 395 | 08/04 03:40 | N19W36 | -26, 39 | 1709 A | 1307 | 36.6 | 38.9 | -2.3 | 1945 | 22.0 | -16.9 | ---- |
| 2011/09/09 11:49 | 340 | 09/06 02:24 | N14W07 | 6, 12 | 513 A | 494 | 88.1 | 81.4 | +6.7 | 521 | 85.3 | +3.9 | 5.2 |
| 2011/09/17 03:05 | 350 | 09/14 00:00 | N22W03 | 10, 9 | 577 B | 534 | 84.0 | 75.1 | +8.9 | 467 | 91.0 | +15.9 | -5.9 |
| 2011/11/12 05:10 | 350 | 11/09 13:36 | N22E44 | 41, -12 | 1366 B | 911 | 54.8 | 63.6 | -8.9 | 1210 | 40.2 | -23.4 | -3.5 |
| 2012/01/22 05:18 | 312 | 01/19 14:25 | N32E22 | 40, 1 | 1153 B | 907 | 54.9 | 62.9 | -8.0 | 674 | 71.3 | +8.4 | ---- |
| 2012/01/24 14:33 | 400 | 01/23 03:38 | N29W20 | -2, 44 | 2002 A | 1645 | 27.3 | 34.9 | -7.6 | 1245 | 38.8 | +3.9 | -0.5 |
| 2012/02/26 21:07 | 350 | 02/24 03:46 | N25E28 | 47, -1 | 779 B | 623 | 75.7 | 65.3 | +10.4 | 678 | 71.0 | +5.7 | -1.0 |
| 2012/03/08 10:53 | 475 | 03/07 01:36 | N17E27 | 47, 1 | 2190[c] B | 1866 | 23.2 | 33.3 | -10.3 | 1402 | 19.0 | -14.5 | -0.8 |
| 2012/03/11 12:52 | 400 | 03/09 04:14 | N17W03 | 17, 31 | 861 B | 822 | 60.3 | 56.6 | +3.7 | 1176 | 41.5 | -15.1 | ---- |
| 2012/03/12 08:45 | 400 | 03/10 17:40 | N17W24 | -4, 52 | 1558 A | 1361 | 34.8 | 39.1 | -4.3 | 1081 | 45.7 | +6.6 | 14.3 |
| 2012/06/16 08:52 | 300 | 06/14 14:36[a] | S17E06 | 33, 20 | 1207 B | 1148 | 42.7 | 42.3 | -0.4 | 1317 | 36.3 | -6.0 | 10.0 |
| 2012/07/14 17:27 | 350 | 07/12 16:49 | S14W01 | 36, 19 | 1548 B | 1502 | 30.7 | 48.6 | -17.9 | 1210 | 40.2 | -8.4 | -5.5 |

[a]The halo CME announcement incorrectly listed the CME onset time as 12:36 UT, [b]STA had a data gap for this event, so STB is used for measurement; [c]New space speed measurements from STEREO/COR2

Column 6 gives $V_{sp}$ along with a suffix A or B indicating which of the STEREO spacecraft data was used for the speed measurement. The speed is the average speed within the COR2 FOV obtained by fitting a straight line to the height-time measurements. Recall that the earthward speed is the primary input to the ESA model. Many solar sources are more than a few degrees from the disk center, so we applied a simple projection correction to the COR2 space speed to obtain the earthward speed (column 7). For example, in the case of the 2012 January 19 CME, the space speed measured in the STB/COR2 FOV was 1153 km/s. The earthward speed ($V_E$) becomes 907 km/s (1153 x cos 32° x cos 22° km/s) since the source location in Earth view was N32E22. When we use this speed in the ESA model (eq. 2), we get a CME travel time of 54.9 h (column 8). This is the travel time of the CME predicted by the ESA model ($t_{ESA}$). Column 9 gives the observed travel time ($t_{obs}$) of the CME as the time elapsed since the first appearance of the CME in the STEREO/COR2 FOV to the onset of the shock at the SOHO spacecraft. For the 2012 January 19 CME, the observed shock transit time is 62.9 h. Thus the predicted travel time is shorter than the actual one by 8 h. The difference between predicted and observed travel times ($\Delta t = t_{ESA} - t_{obs}$) is listed in column 10. The deviation $\Delta t<0$ ($\Delta t>0$) indicates CME arrival later (earlier) than predicted, and represents the prediction error of the ESA model.

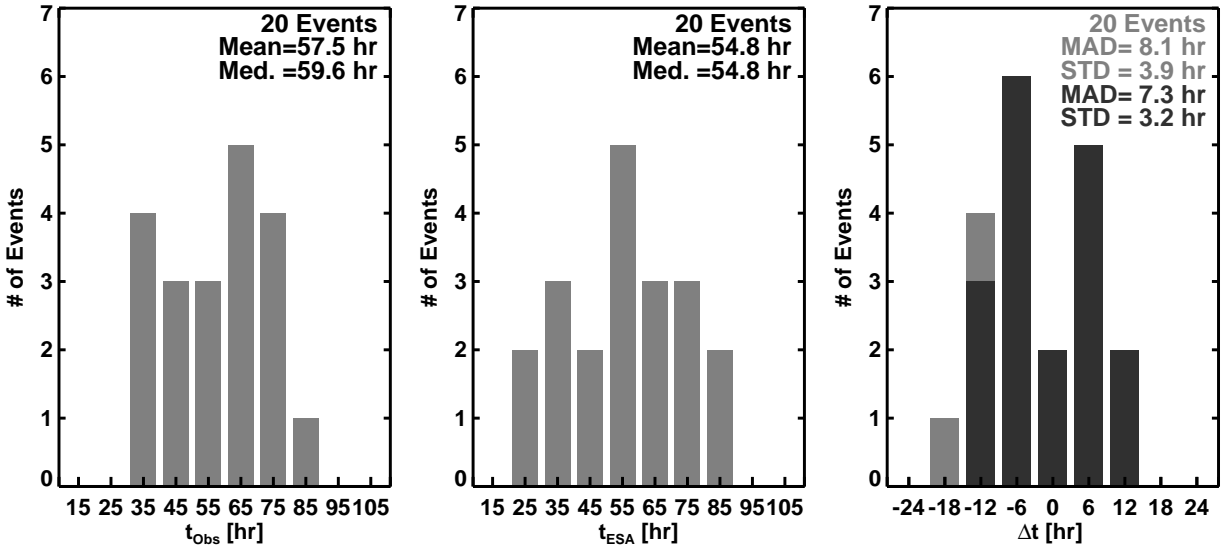

*Figure. 1 (a) The observed ($t_{obs}$) and (b) predicted ($t_{ESA}$) Sun-Earth travel times for the 20 shocks listed in Table1. (c) The deviation $\Delta t$ of $t_{obs}$ from $t_{ESA}$. The mean and median values of the distributions are given on the plots. The two outlier events in (c) shown in gray are discussed in the text.*

Figure 1 compares $t_{obs}$ and $t_{ESA}$ for the 20 events in Table 1 and shows the distribution of $\Delta t$. The range of $t_{obs}$ is from 33 to 81 h, with a mean value of 57.5 h. The range of $t_{ESA}$ is from 23 to 88 h with a mean of 54.8 h. Thus the predicted travel time range is very similar to the observed one. The prediction error $\Delta t$ is in the range -17.9 h to +10.4 h. However, there are only 2 events outside ±12 h. Thus the ESA model is able to predict the shock arrival within ±12 h in 90% of cases considered in this paper. The mean absolute deviation (MAD) is 8.1 h for all events and 7.3 h when the two large-$\Delta t$ events are excluded. The corresponding standard deviations are 3.9 and 3.2h, respectively. Both the large-deviation events have $\Delta t <0$ (-13.4 h for the 2011 February 18 shock and -17.9 h for the 2012 July 12 shock, which are discussed in the next section). Finally, the root mean square (RMS) error is 9.1 h.

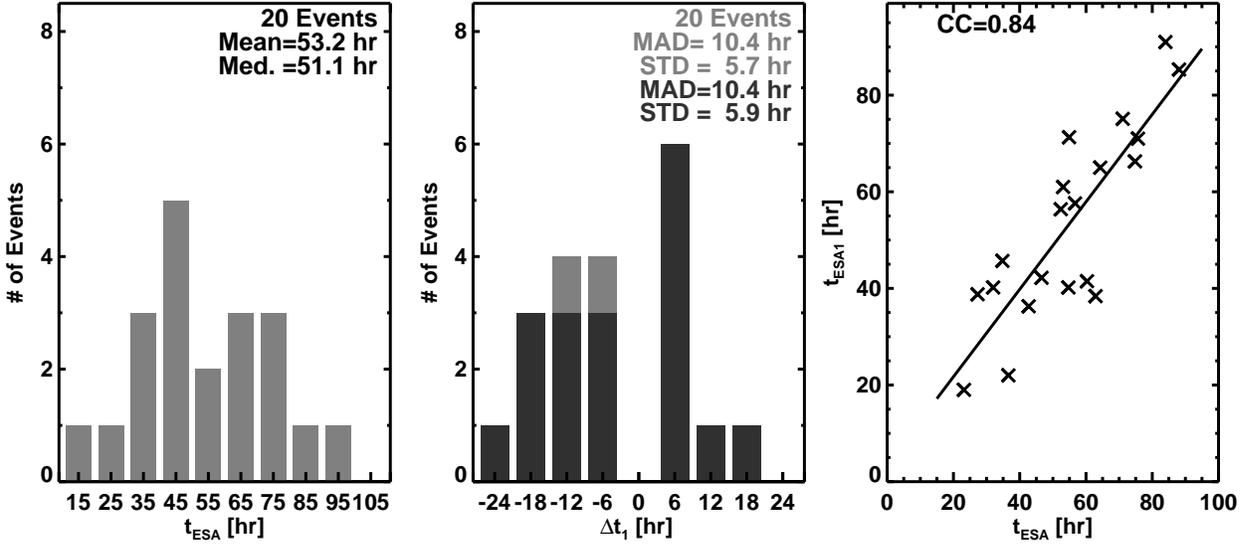

*Figure 2. (a) CME travel time predicted by the ESA model ($t_{ESA1}$) for the measured ecliptic speed toward Earth ($V_{E1}$). (b) The deviation of the observed travel time from the predicted travel time ($\Delta t_1 = t_{ESA1} - t_{obs}$). The two events distinguished by gray are the outliers noted in Fig. 1. (c) Correlation between $t_{ESA1}$ and $t_{ESA}$ for the 20 events in Table 1. The probability that the high correlation is by chance is $4.6 \times 10^{-7}$. The correlation also validates the use of earthward speed either by deprojection or by ecliptic measurement.*

*Taktakishvili et al.* [2009] considered a smaller sample (14 events vs. 20 in the present work) and found a mean travel-time error of 5.9 h for the ENLIL model. This is only 27% better than the value obtained in the present work. When we exclude the two outliers in *Taktakishvili et al.* [2009], their mean absolute error becomes 5.1 h, which is 30% better than our value (7.3 h) with the two outliers excluded. *Taktakishvili et al.* also used the ESA model with cone-model speeds and obtained a mean error of 8.4 h, which is similar to our values, and slightly worse if the outliers are excluded. *Xie et al.* [2006] used cone models to correct the sky-plane speed for projection effects for a sample of 40 events and found a mean error of the ESA model to be 7.8 h, which is similar to the error obtained in the present work. The present work, therefore, validates the cone model method of correcting for the projection effects. It is also worth noting that the initial tests of the "drag based model" of Vrsnak et al. (2013) revealed that the typical mean absolute error in predicting the CME arrival (not the shock arrival) is ~12 h.

In order to check the validity of the earthward speed obtained by simple deprojection, we measured the CME speeds in the ecliptic plane using STEREO data. Essentially we made the CME height-time measurements at position angles 90° (STA) and 270° (STB), neglecting the solar B0 angle (the heliographic latitude of the ecliptic). The measured speed is the earthward speed ($V_{E1}$), which is listed in column 11 of Table 1. Using $V_{E1}$ as input in eq. (2), we obtained the predicted travel time ($t_{ESA1}$) and the deviation ($\Delta t_1 = t_{ESA1} - t_{obs}$) as listed in columns 12 and 13, respectively. Figure 2 shows that $t_{ESA1}$ is distributed similar to $t_{ESA}$: the mean and median values are 53.2 and 51.1 h, respectively. The error $\Delta t_1$ is distributed slightly broader than $\Delta t$ with a mean value of 10.4 h. The prediction error is within ±12 h only for 70% of the cases (compared to 90% from the deprojection method). The outliers are also different, except for the 2011 February 11 shock. Figure 2c shows a scatter plot between $\Delta t$ and $\Delta t_1$. In general there is a good agreement between the two (the correlation coefficient is 0.84), confirming the importance of estimating the earthward speed as closely as possible. One of the main sources of error in using the ecliptic measurements is that there may be projection effects and deflections in the longitudinal direction.

## 3. Outlier Events

As we noted above, there are two outliers events with large negative $\Delta t$ (< -12 h) that can be readily recognized as cases with significant propagation effects, either due to CME interaction [*Gopalswamy et al.*, 2001; *Manoharan et al.* 2004] or due to coronal-hole deflection [*Gopalswamy et al.* 2009a; *Xie et al.*, 2013; *Mäkelä et al.*, 2013]. The propagation effects essentially decrease the effective earthward speed of the CMEs, so they take longer than predicted to arrive at Earth and hence make $\Delta t$ negative. In an operational setting, recognizing the environment of an eruption can significantly improve the prediction capability by the expectation of a delayed arrival of the CME. These events are examined in more detail in this section.

### 3.1 The 2011 February 15 CME

The shock driven by this CME arrived about 13.4 h after the predicted time ($\Delta t$ = -13.4 h). The CME originated from S12W18 on 2011 February 15 at 02:36 UT and was observed as a full halo in the LASCO FOV. The CME was observed in quadrature by STA and STB as a limb event [*Gopalswamy et al.* 2012]). The expansion speed of a CME is defined as the rate at which the lateral extent of the CME increases [see e.g., *dal Lago et al.* 2003; *Schwenn et al.* 2005]. Thus the sky-plane speed of a full halo CME corresponds to half the expansion speed. For the 2011 February 15 CME, the sky-plane speed from SOHO/LASCO gives the expansion speed while that in STA and STB correspond to the radial speed, in agreement with the full ice-cream cone model [*Gopalswamy et al.* 2009b; 2012]. STEREO movies made from COR1 and COR2 images show that the CME in question was preceded by several CMEs from the same region. We were able to count about 11 preceding CMEs starting from the partial halo CME on 2011 February 13 at 18:36 UT from roughly the same position angle as the 02:36 UT CME. Figure 3 shows three of the preceding CMEs in STB view along with the 02:36 UT CME (marked as CME4). The effect of the slower preceding CMEs is to increase the effective drag on CME4. The drag force is given by [*Cargill*, 2004],

$$F_d = -C_d A \rho (V-V_{sw})^2, \qquad (3)$$

where $C_d$ is the drag coefficient, $A$ is the cross-sectional area of CME4, $\rho$ is the ambient density, and $V$ is the speed of CME4, and $V_{sw}$ is the ambient solar wind speed. In this case the $V_{sw}$ is replaced by the speed of the preceding CMEs and the ambient density is replaced by the density of preceding CMEs. Since the CME density is significantly enhanced with respect to the ambient medium, it is likely that the drag on CME4 increases significantly and depends on how many CMEs it crosses or sweeps on the way. This additional drag is likely to have caused the extra travel time, thus deviating significantly from the predicted travel time [see also *Manoharan et al.* 2004; *Temmer et al.* 2012].

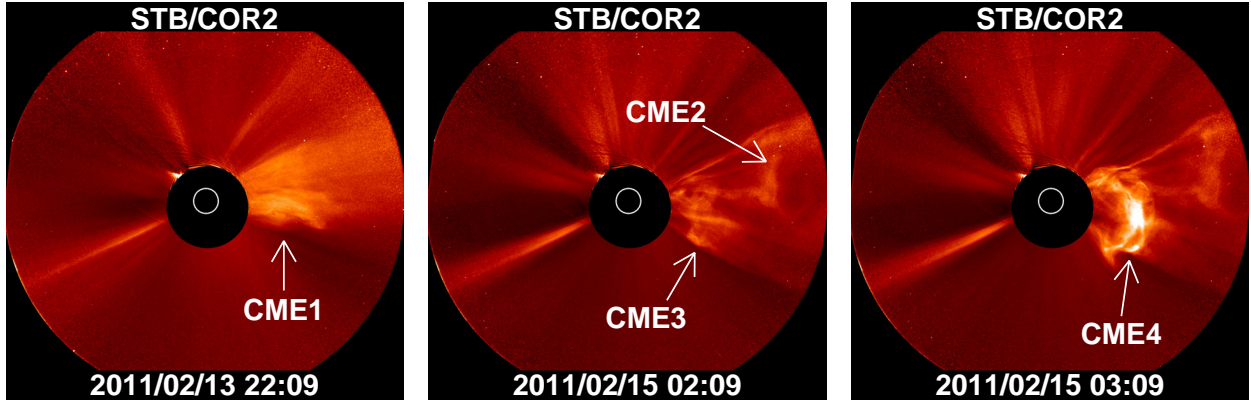

*Figure 3. A series of snapshot images showing 3 CMEs (CME1, CME2, and CME3) preceding the 2011 February 15 CME (CME4) that was driving the 2011 February 18 shock. CME2 and CME3 were observed within the preceding 4 hours. There were 11 CMEs between CME1 and CME4, mostly from the same source region. All these CMEs were in the path of CME4 en route to Earth.*

### 3.2 The 2012 July 12 CME

The arrival time of the shock driven by the 2012 July 12 CME showed the largest deviation from the ESA model prediction: $\Delta t$ = -17.9 h. The CME originated in the northern part (S14W01) of NOAA AR 11520 located at the central meridian. If the CME propagates radially out, we expect the central position angle (CPA) to be $180^o$. On the other hand the fastest moving section of the CME in the LASCO/C2 FOV was at $232^o$, indicating a deviation of ~$42^o$ from the radial direction. This indicates a westward deflection of ~$42^o$ from the Sun-Earth line. In STB view, the CME was located just $19^o$ behind the west limb. The CPA in STB/COR1 FOV was $241^o$. The CME shape in Fig. 4(a) indicates a northward deflection. The reason for the west – northwest deflection of the CME can be readily inferred from the coronal holes situated to the eastern side of the source active region (Fig. 4c,d). Deflections as large as $57^o$ were inferred from the source regions of "driverless" shocks [*Gopalswamy et al.* 2010b]. In the present event, we infer a deflection of ~$52^o$, which is within the range of deflection angles in *Gopalswamy et al*. [2010b]. By measuring the area ($A$), the average photospheric magnetic field strength ($B$) within the coronal holes, and the distance ($r$) between the coronal-hole centroid and the CME source region we obtain the coronal hole influence parameter (CHIP) as $AB/r^2$, which is the magnitude of the force exerted by the coronal hole on the CME. CHIP =1 G for the 2012 July 12 event, with the main contribution (0.8 G) coming from the large coronal hole located to the east (S09E22) of the eruption region. The coronal hole influence is directed along position angle 270°, consistent with the west-northwest deflection observed in the images.

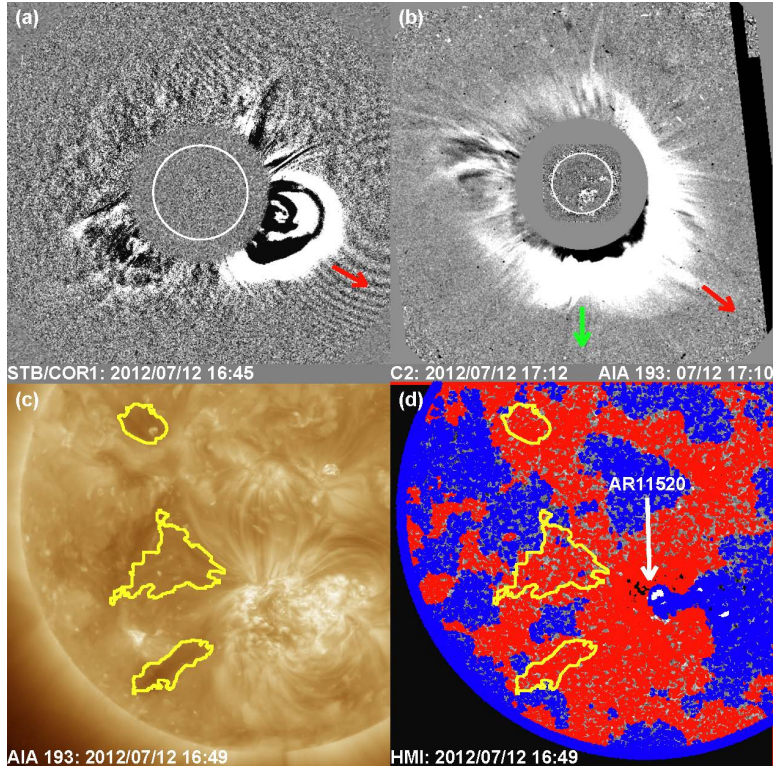

*Figure 4. Non-radial motion of the CME due to deflection by coronal holes. (a) STB/COR1 image showing the CME at the west limb at 16:45 UT on 2012 July 12. (b) The CME in the LASCO/C2 FOV with the expected CPA (180°, green arrow) and the actual CPA (232°, red arrow) marked. (c) The deepest sections of the coronal hole located on the eastern side of the source active region in the Atmospheric Imaging Assembly (AIA) image. The contours correspond to regions below 50% of the median intensity of the solar disk in EUV. (d) A magnetogram from SDO's Helio Magnetic Imager (HMI) showing the eruption region (AR 11520 pointed by an arrow) and the magnetic polarity (red is negative and blue is positive) of the coronal-hole sections. The centroids of the three coronal holes (top to bottom) are: N24E33, S09E22, and S36E29. The CHIP and FPA are: (0.11 G, 222°), (0.77 G, 261°), and (0.27 G, 314°). The net value of CHIP is 1.0 G acting along PA = 270°. The coronal hole at S09E22 has the most influence on the CME.*

## 4. Discussion

The ESA model, derived originally from quadrature observations, could not be tested for about 12 years due to lack of similar observations. The opportunity finally arose when SOHO and STEREO were observing in quadrature during 2010-2012. SOHO played the role of the *Helios* spacecraft in providing the *in-situ* observation of shocks. The STEREO coronagraphs played the role of the Solwind coronagraph in remote-sensing CMEs still near the Sun. In addition, SOHO also remotely sensed the CMEs as halo events, which was useful in pairing CMEs with the corresponding shocks.

Predicting the shock arrival with an error of 7.3±3.2 h is certainly at the level of state-of-the-art numerical simulation models. We already made a comparison with the ENLIL model prediction as reported by *Taktakishvili et al*. [2012]. NOAA Space Weather Prediction Center (SWPC) has been using the ENLIL model over the past few years in an operational setting and has compiled error estimates for 42 events [*Biesecker*, 2013, private communication]. The mean absolute and RMS errors for these 42 events were found to be 7.3 and 9.3 h, respectively. These are virtually the same as the corresponding values (7.3 and 9.1 h) for the 20 events in Table 1. When we compared our list in Table 1 with the SWPC list, we found that 15 events are common to the two lists. The error $\Delta t_2$ in these events (deviation of the ENLIL model travel time from the observed travel time) is listed in the last column of Table 1. The error is in the range from -9.8 to +14.3 h, with the mean absolute and RMS errors being 4.6 and 6.2 h, respectively. For these 15 events, the corresponding ESA model errors are 8.5 and 9.4 h, respectively indicating that the ENLIL model is better by 34 to 46%. Unfortunately, the sample size is significantly reduced when we consider events overlapping between the two lists. It must be noted that the ENLIL model errors computed by *Taktakishvili et al*. [2009] correspond to post-prediction analysis, while the *Biesecker* [2013] ones, by their operational nature, are true predictions and are made with incomplete and less-than ideal input data. Currently, the CMEs are input as uniform density spheres to the ENLIL model, so they are not realistic. The ENLIL model prediction is heavily dependent on these inputs, so better-defined inputs are likely to result in better predictions.

One of the important outcomes of this work is that the presence of coronal holes near the eruption region and preceding CMEs might significantly affect the shock travel time. Our data are consistent with the picture that coronal holes deflect CMEs towards and away from the Sun-Earth line, thus altering the earthward speed and hence the travel time. In the preceding section, we explained the large deviation of the 2012 July 23 event in terms of CME deflection by coronal holes. In order check whether other events in Table 1 were affected by coronal holes, we looked for EUV coronal holes on the disk around the time of the eruptions. There were significant coronal holes only in 6 cases apart from the 2012 July 12 event. Thus the influence of coronal holes is negligible for the rest of the 13 events. We computed the CHIP values and the direction in which the coronal hole influence acts for the coronal holes for each eruption. The position angle (PA) along which the net force acts (FPA) is listed in Table 2 along with the CHIP and the PA along which CME height-time measurements are made (MPA). The difference PA (DPA) between MPA and FPA is also listed in Table 1. The $\Delta t$ values (extracted from Table 1) are within the MAD (8.1 h) for four events. The three events with $\Delta t$ well above the MAD are the two outlier events discussed in section 3 and the 2012 February 24 event. While the outliers have negative $\Delta t$, the 2012 February 24 event has a large positive $\Delta t$. The CHIP value (1.5 G) for the 2012 February 24 event is the highest in Table 2 and the FPA is to the northwest (311°). It appears that the earthward speed of this CME was higher than our estimate because the shock arrival is ~10.4 h ahead of the prediction. This can be understood in terms of coronal deflection towards the Sun-Earth line (opposite of what happened to the 2012 July 12 CME). The source location (N25E28) was between the coronal hole (N12E42) and the disk center, so it is likely that the CME was deflected toward the Sun-Earth line. A second coronal hole to the north of the source region (N41W01) might have also played a secondary role in doing the same in the north-south direction albeit with a smaller CHIP (0.6 G).

*Table 2. List of events with coronal holes visible on the disk.*

| CME Date | CHIP (G)$^a$ | FPA (deg)$^b$ | MPA (deg)$^c$ | DPA (deg)$^d$ | $\Delta t$ (h) | Source | CH Loc. |
|---|---|---|---|---|---|---|---|
| 2010/08/01 | 0.06 | 134 | 84 | 50 | +6.4 | N20E36 | S67E00 |
| 2011/02/15 | 0.12 | 340 | 189 | 151 | -13.4 | S12W18 | S59W15 |
| 2011/06/21 | 1.4 | 358 | 65 | 67 | +6.1 | N16W08 | S17W09 |
| 2011/09/06 | 0.57 | 272 | 70 | 158 | +6.7 | N14W07 | N10E40 |
| 2012/01/23 | 0.33 | 281 | 326 | 45 | -7.6 | N29W20 | N42E22 |
| 2012/02/24 | 1.5 | 311 | 1 | 50 | +10.4 | N25E28 | N12E42$^e$ |
| 2012/07/12 | 1.0 | 270 | 180 | 90 | -17.9 | S14W01 | S09E22 |

$^a$The coronal hole influence parameter (the net force extorted by the coronal holes on the CME
$^b$Forcing position angle, the position angle along which the net force due to the coronal holes act
$^c$The PA along which the CME height-time measurements are made (measurement PA)
$^d$The difference between MPA and FPA
$^e$A second coronal hole at N41W01 had CHIP = 0.6 G

The 2011 June 21 CME also has a high CHIP (1.4 G) and the *Δt* value is positive (+6.1 h). Even though the source location was in the northwest quadrant, the main body of the CME headed in the NE direction (see
http://cdaw.gsfc.nasa.gov/CME_list/UNIVERSAL/2011_06/jsmovies/2011_06/20110621.03161 0.p065g/c2_rdif.html), probably due to the coronal hole. The Earthward speed seems to be slightly larger, somewhat similar to the 2012 February 24 event, but to a smaller extent. The CHIP value is rather small (0.12 G) for the 2011 February 15 CME, which means the coronal hole deflection may not be the major reason for the large deviation. This is consistent with the explanation provided in terms of CME interaction for this event (see section 3.1). Thus the coronal hole influence can delay or speed up shock arrival depending on the position of the coronal hole with respect to the eruption region and the Sun-Earth line [see *Gopalswamy et al.* 2009b for more details].

In section 3.1, we suggested that preceding CMEs might increase the effective drag force, thereby slowing down a CME and hence delaying the 1-AU arrival time. We specifically discussed the 2011 February 15 CME, in which the shock arrived significantly late compared to both ESA and ENLIL predictions (see Fig. 3 and Table 1). We now discuss the preceding CMEs in other events in Table 1.   Table 3 shows the number of preceding CMEs that were identified in STEREO images within an interval of 24 h before the onset times of the ESA test events in Table 1. The preceding CMEs (PCMEs) appear to travel roughly along the same position angle as that of the CMEs in Table 1.

*Table 3. The number of preceding CMEs within a 24 h interval preceding the CMEs in Table 1*

| CME Date and Time | #PCMEs (W≥30º) | #PCMEs (W<30º) | Total (LASCO) | $\Delta t$ (h) |
|---|---|---|---|---|
| 2010/02/12 13:31 | 1 | 0 | 1 (0)$^a$ | -11.7 |
| 2010/04/08 04:30 | 0 | 0 | 0 (0) | -8.7 |
| 2010/08/01 08:24 | 1 | 0 | 1 (0) | +6.4 |
| 2011/02/15 02:36 | 4 | 3 | 7 (3) | -13.4 |
| 2011/03/07 14:48 | 0 | 3 | 3 (0) | +11..8 |
| 2011/06/21 03:16 | 0 | 1 | 0 (0) | +6.1 |
| 2011/08/02 06:36 | 1 | 1 | 2 (2) | -10.2 |
| 2011/08/03 13:17 | 3 | 0 | 3 (2) | -5.5 |
| 2011/08/04 03:40 | 3 | 0 | 3 (3) | -2.3 |
| 2011/09/06 02:24 | 0 | 1 | 1 (1) | +6.7 |
| 2011/09/14 00:00 | 0 | 1 | 1 (1) | +8.9 |
| 2011/11/09 13:36 | 0 | 1 | 1 (1) | -8.9 |
| 2012/01/19 14:25 | 2 | 0 | 2 (2) | -8.0 |
| 2012/01/23 03:38 | 1 | 0 | 1 (1) | -7.6 |
| 2012/02/24 03:46 | 1 | 0 | 1 (1) | +10.4 |
| 2012/03/07 01:36 | 2 | 0 | 2 (0) | -10.3 |
| 2012/03/09 04:14 | 0 | 0 | 0 (0) | +3.7 |
| 2012/03/10 17:40 | 2 | 0 | 2 (2) | -4.3 |
| 2012/06/14 14:36 | 2 | 1 | 3 (2) | +0.4 |
| 2012/07/12 16:49 | 0 | 0 | 0 (0) | -17.9 |

$^a$LASCO data gap during 01:31 – 13:31 UT on 2010 February 12.

There were 33 PCMEs in all, of which 23 had width (W) ≥30º. The remaining 10 PCMEs were narrow (W<30º).  The PCMEs detected by both STEREO and LASCO were all wider CMEs (width ≥30º). Note that about 36% the PCMEs were not detected by LASCO, probably because they were narrow and did not expand enough to show up above the LASCO occulting disk.  The 2011 February 15 CME stands out in that it had seven PCMEs (We had noted 11 PCMEs in section 3.1, but it was over a larger interval).  The maximum number of PCMEs is only 3 in other events.  Table 3 shows that most of the events with at least one preceding wide CME have negative $\Delta t$, consistent with the delayed arrival of the associated shocks in Table1. The 2012 March 07 CME was preceded by another superfast CME (~2600 km/s) that was ejected just an hour earlier from the same active region. The preceding CME was to the north of the Sun-Earth line, so did not deliver a significant impact on Earth. However, the CME in question was much closer to the Sun-Earth line and traveled through the flank of the preceding CME. The shock arrived ~10.3 h after the predicted time, likely due to this interaction. The 2012 February 24 event is a notable exception because the shock arrival is ahead of the model prediction by 10.4 h. However, we have shown in Table 2 that this CME was deflected towards the Sun-Earth line by coronal holes.

The ENLIL model run for the 2011 February 15 CME [*Biesecker*, 2013, private communication] used only two PCMEs: the halo CME on 2011 February 14 at 18:24 UT (CME3 in Fig.3 and a previous partial halo CME on 2011 February 13 CME at 18:36 UT (CME1 in Fig. 3). Note that CME1 was just outside the 24-hour window. We also note that the quadrature observations helped identify many PCMEs not detected by LASCO. On the other hand these PCMEs were limb events for the STEREO coronagraphs and hence were readily detected.

While the presence of coronal holes near the eruption region can be readily seen from inner coronal images in EUV or X-rays, it is difficult to identify the preceding CMEs heading toward Earth (especially the narrow ones) using coronagraphs observing from the Sun-Earth line. However, a coronagraph located at Sun-Earth L5 or L4 can observe the preceding CMEs and also measure the earthward speed of CMEs [*Gopalswamy et al*. 2011a,b]. The quadrature observations presented here point to the importance of coronagraphic observations of Earth-directed CMEs away from the Sun-Earth line. A coronagraph in a solar polar orbit can also provide such observations. *A priori* knowledge on the existence of coronal holes and nonradial CME motion helps make a better CME travel time prediction using the ESA model. Coronagraphs with extended FOVs are very useful in identifying the non-radial motion. One may also have to consider CME interaction with other large-scale structures such as the heliospheric current sheet and high-speed streams. Correctly modeling the CME location relative to high speed streams and the current sheet is also clearly important in making better travel-time prediction of shocks.

The ESA model assumes a constant solar wind speed of ~406 km/s (see eq. (1)). From Table 1, we see that the solar wind speed ranges from 300 km/s to 600 km/s, with an average value of 372 km/s. We compiled this information to see if the ambient flow speed affects the CME travel time. According to eq. (1), lower solar wind speed $u_c$ implies more drag that would result in a longer travel time than the $u_c$ = 406 km/s the ESA model uses. In fact there are clearly more $\Delta t < 0$ events in Table 1 (13 vs. 7), which is consistent with the lower solar wind speed. The largest deviation in the solar wind speed from 406 km/s is for the 2011 June 23 shock: the ambient solar wind was speed of 600 km/s and the shock arrival was earlier than predicted by ~6 h.

Finally, a comment on the definition of quadrature is in order. Recall that we allowed a maximum deviation of ~30° in defining quadrature between two observing spacecraft. In fact the deviation was ~30° in only one event (the first event in Table 1). STA was in quadrature with LASCO (deviation was only 14°), but a data gap prevented us from using the STA data. Using STB data means the speed might be underestimated by ~15%, so the correct space speed and Earthward speed are likely to be 997 km/s and 880 km/s, respectively. The new travel time is 56.6 h and the deviation becomes larger: -19.4 h. As we had shown in Table 3, this event was preceded by a wide CME from the same source region, so the delay is consistent with what is expected from interacting events. In fact, this event was almost outside the ±12 h range, but now definitely goes outside this range. If we assume that we systematically underestimate CME space speeds by a factor 1/cos(angle-to-limb), then MAD = 9.2 h (8.1 h) for all (two outliers excluded). With this correction, the 2010 February 12 event also becomes an outlier, excluding which we get MAD = 7.5 h, not significantly different from the original 7.4 h. There is no significant change in the prediction values because the deviation from exact quadrature is not significant except for two events: the deviation is ≤6% in 16 cases, and one each with 7%, 9%, 11% and 15%.

**5. Conclusions**

Making use of the opportunity for quadrature observations from STEREO and SOHO, we tested the capability of the empirical shock arrival model for a set of 20 events. Limiting the study to full halo CMEs that had speeds exceeding 450 km/s, we found that the ESA model predicts shock arrival with a mean absolute error of 7.3±3.2 h. The RMS error of the ESA model is 9.1 h. Other results of the study can be summarized as follows.

1. The performance of the ESA model is comparable to the ENLIL model, whose mean absolute and RMS errors (7.3 and 9.3 h) are virtually the same as those of the ESA model (7.3 and 9.1 h). However, for a smaller sample of 15 events, the ENLIL model predicts 34-46% better.
2. Accurate estimation of earthward speed is needed to minimize the prediction error. CME-CME interaction and CME-coronal hole interaction are two significant effects that can lead to large deviation from the ESA model.
3. Observations made from vantage points off the Sun-Earth line (in the ecliptic or above the ecliptic) can provide a better estimate of the speed of Earth-directed CMEs and identify preceding CMEs that can significantly alter the CME travel time.
4. Coronal holes observed in EUV/X-ray images and the underlying photospheric magnetic field are needed to provide information on the potential deviation of CME trajectories and hence their travel time to Earth.

**Acknowledgments:** We thank D. Webb and D. Biesecker for reading the manuscript and making helpful suggestions. We also thank S. Akiyama for help with Fig. 4. We thank the SOHO, Wind and ACE science teams for making the shock data available on line. STEREO is a mission in NASA's Solar Terrestrial Probes program. SOHO is a project of international cooperation between ESA and NASA. This research was supported by NASA LWS TR&T program.